\documentclass[aps,superscriptaddress,showpacs,
  twocolumn,nofootinbib, showkeys, amsmath, amssymb]{revtex4}

\usepackage{natbib}
\usepackage{graphicx}
\usepackage{hyperref}
\usepackage{amsfonts}
\usepackage{amssymb}
\usepackage{amsmath}
\usepackage{tikz}
\usepackage{url}

\newcommand{\ddr}[1]{
  \frac{\mathrm{d} #1}{\mathrm{d}r}}
\newcommand{\R}{
  \mathbb{R}}
\newcommand{\floor}[1]{
  \left\lfloor #1 \right\rfloor}

\begin{document}

\title{Competition for Popularity in Bipartite Networks}

\author{Mariano Beguerisse D\'iaz}
\email{m.beguerisse-diaz08@imperial.ac.uk}
\affiliation{Centre for Integrative Systems Biology,
  Imperial College London, South Kensington Campus, London, SW7 2AZ, U.K.}
\affiliation{Division of Biology,
  Imperial College London, South Kensington Campus, London, SW7 2AZ, U.K.}
\author{Mason A. Porter}
\email{porterm@maths.ox.ac.uk}
\affiliation{Oxford Centre for Industrial and Applied Mathematics,
  Mathematical Institute, University of Oxford, OX1 3LB, U.K.}
\affiliation{CABDyN Complexity Centre, 
  University of Oxford, OX1 1HP, U.K.}
\author{Jukka-Pekka Onnela}
\email{onnela@hcp.med.harvard.edu}
\affiliation{CABDyN Complexity Centre, 
  University of Oxford, OX1 1HP, U.K.}
\affiliation{Department of Physics, University of Oxford, OX1 3PU, U.K.}
\affiliation{Department of Biomedical Engineering and Computational Science, 
  Helsinki University of Technology, FIN-02015 HUT, Finland}
\affiliation{Harvard Kennedy School, Harvard University, Cambridge, 
  MA 02138, U.S.A.}


\begin{abstract}
 
We present a dynamical model for rewiring and attachment in bipartite networks in 
which edges are added between nodes that belong to catalogs that can either be fixed 
in size or growing in size.  The model is motivated by an empirical study of data from 
the video rental service Netflix, which invites its users to give ratings to the 
videos available in its catalog. We find that the distribution of the number of 
ratings given by users and that of the number of ratings received by videos both
 follow a power law with an exponential cutoff.  We also examine the activity patterns 
of Netflix users and find bursts of intense video-rating activity followed by long 
periods of inactivity.  We derive ordinary differential equations to model the 
acquisition of edges by the nodes over time and obtain the corresponding 
time-dependent degree distributions.  We then compare our results with the Netflix 
data and find good agreement.  We conclude with a discussion of how catalog models 
can be used to study systems in which agents are forced to choose, rate, or 
prioritize their interactions from a very large set of options.

\end{abstract}

\pacs{89.75.Hc, 89.65.-s, 05.90.+m}

\keywords{Bipartite Networks, Human Dynamics, Catalog Networks, Bursts, Rate Equations}

\maketitle

\vspace{.1 in}


\begin{bfseries} 

\textit{Human dynamics}, which is concerned with the characterization of human 
activity in time, has been the  subject of intense and exciting research 
over the last few years~\cite{barabasi-2005-435,evans-2008-3,J.P.Onnela05012007}. 
In one typical 
problem setting, individuals are endowed with limited resources, and there 
are numerous activities, behaviors, and/or products that compete against 
each other for those resources.  Although such situations admit a natural 
formulation using bipartite (two-mode) networks that connect individuals 
to activities, human dynamics has surprisingly seldom been studied from 
this perspective.  In the present paper, we analyze bipartite networks 
constructed from a large data set of video ratings by the users of a video 
rental company over a period of six years. To analyze the time evolution of 
these networks, we introduce the concept of a \textit{catalog network}, 
and we use this approach to explore the driving forces behind the video 
rating behavior of individuals.  We believe that such a framework can be used to study
 many other phenomena in human dynamics that involve the 
allocation of and competition for scarce resources.

\end{bfseries}    


\section{Introduction}

Numerous natural and man-made systems involve interactions between large
numbers of entities. The structural configuration of interactions is 
typically rather complicated, so the study of such systems often benefits 
greatly from network representations \cite{albert-2002-74,Newman:2003,guido}.
A network is usually abstracted mathematically as a graph whose nodes 
represent the entities and whose edges represent the interactions between the 
entities \cite{HandbookGraphTheory}.  In many cases, edges can be weighted 
or directed, and more complicated frameworks such as hypergraphs can 
also be employed. The number of edges connected to a node in an unweighted network 
is known as its \textit{degree}, and the \textit{degree distribution} of 
a network is given by the collection of numbers that give the fraction of 
nodes that have degree $k$ (for all values of $k$) \cite{Newman:2003}.  
In weighted networks, one considers the weight of an 
edge rather than simply whether or not it exists.

Because networked systems are not static, the last decade has witnessed a 
particular interest in models that attempt to address their growth and 
evolution \cite{guido}.  Perhaps the best-known model of network growth 
was formulated by Barab\'asi and Albert 
\cite{BarabasiAlbert1999,albert-2002-74}.  Similar models were also 
constructed decades earlier by Simon \cite{HERBERTA.SIMON12011955} and 
Price \cite{Price:1965}. Barab\'asi and Albert examined networks arising 
from diverse settings and found that their degree distributions often seemed to 
follow power laws, which are functions of the form 
$f(x) \sim x^{-\alpha}$ (with $\alpha > 0$).  They proposed a growth mechanism, 
which they called \textit{preferential attachment} (Price had called it 
\textit{cumulative advantage}) to try to explain their observations. One starts 
with a small seed network and---in the simplest form of the mechanism---iteratively 
adds individual nodes that each possess exactly one edge.  One connects each new 
node to an existing one chosen at random with probability proportional to its degree.  
That is, the probability to choose node $m_i$ with degree $k_i$ is 
\begin{equation}
 	P(m_i) = \frac{k_i}{\sum_{j=1}^N k_j}\,, 
        \nonumber
\end{equation}
where the total number of nodes $N$ indicates the size of the network.  
Because nodes with higher degrees have correspondingly higher probabilities 
to receive new edges, the preferential attachment growth mechanism leads 
naturally to a power-law degree distribution 
\cite{BarabasiAlbert1999,PhysRevLett.85.4629}.

\begin{figure}
  \begin{center}
     \includegraphics[width=170px]{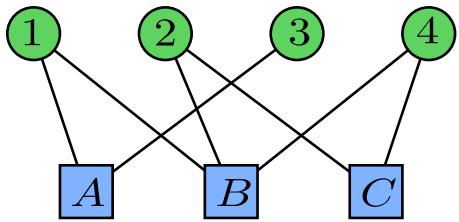}
  \end{center}
  \caption{(Color online) A bipartite network with nodes in the partite sets 
           $\mathcal{U} = \{1, 2, 3, 4\}$ and $\mathcal{M} = \{A, B, C\}$.  
           Each edge connects a number to a letter.}
   \label{bipNetworkGraph}
\end{figure}

Because of ideas like preferential attachment and the resulting insights 
on the origin of heavy-tailed degree distributions that one sees, e.g., 
in the World Wide Web or scientific collaboration networks, the study of networks 
has grown immensely during the last ten years \cite{Newman:2003, NewmanSciCol, guido}.
However, most of this research 
has concentrated on one-mode (unipartite) networks, in which all of the nodes are 
of the same type.  It is perhaps under-appreciated that other graph structures are 
also very important in many applications \cite{Latapy200831}. Even the 
simplest generalization, known as a two-mode or \textit{bipartite network}, 
has been studied much more sparingly than unipartite networks.  Bipartite 
networks contain two categories (partite sets) of nodes: 
$\mathcal{U} = \{u_{1}, u_{2}, \dots , u_{U}\}$ (with $U$ members) and 
$\mathcal{M} = \{m_{1}, m_{2}, \dots, m_{M}\}$ (with $M$ members).  
As shown in Fig.~\ref{bipNetworkGraph}, each (undirected) edge connects a 
node in $\mathcal{U}$ to one in $\mathcal{M}$ \cite{HandbookGraphTheory}.  
Bipartite networks abound in applications: They can represent affiliation 
networks in which people are connected to organizations or committees 
\cite{MasonPorter05172005}, ecological networks with links between 
cooperating species in an ecosystem \cite{Saavedra07532}, and more 
\cite{Zhang20086869, Guillaume04bipartitegraphs, PhysRevE.72.036120,
 evans:056101,baseball}.

A bipartite network possesses a degree distribution
for each of the two node types.  We denote the adjacency matrix of a weighted
 bipartite network by $\mathbf{G} \in \mathbb{R}^{U \times M}$.  
Each matrix element $\mathbf{G}_{ij}$ has a nonzero value if and only if an edge exists 
between nodes $u_i$ and $m_j$. 
We denote the matrices that result from the two unipartite projections as 
$\mathbf{G}_{\mathcal{U}} = \mathbf{G}\mathbf{G}^T \in \R^{U \times U}$
and $\mathbf{G}_{\mathcal{M}} = \mathbf{G}^T\mathbf{G} \in \R^{M \times M}$. 
The degree of a node in a unipartite projection network is then 
the number of nodes of the same type with which the node shares at least one neighbor 
in the original bipartite network.  The node strengths similarly incorporate 
connection strengths from the 
original bipartite network.  (Recall that the ``strength" of a node is the sum of 
the strengths of the edges connected to it.)  For example, in an unweighted 
affiliation network, the two projections give the weighted connection strength 
(the number of common affiliations) 
among individuals and the interlock (the number of common people) among organizations 
\cite{ceo,MasonPorter05172005}.

Many of the real-life systems that can be represented by bipartite networks are 
dynamic,  as the existence and connectivity of both nodes and edges can change 
in time. For example, a person might retire or leave one organization to join 
another.  One of the simplest types of changes is edge rewiring, in which one end 
of an edge is fixed to a node and the other end moves from one node to another 
(such as in the aforementioned change of affiliation).  Because of the important 
insights they can offer, network rewiring models have received increasing attention 
\cite{WattsStrogatz,PhysRevE.72.036120,evans:056101,Dorogovtsev2003396,
PhysRevE.72.026131,fan026103,lind}.  They are closely related to abstract urn models 
from probability theory \cite{polya,feller,Godreche0953}, models of 
language competition \cite{Stauffer2007835}, and models of transmission of cultural 
artifacts \cite{Bentley2003}.  More generally, they can help lead to a better 
understanding of any system in which the nature or existence of an interaction among 
agents changes over time \cite{evans-2008-3}.

The rest of our presentation is organized as follows.  In Section \ref{SecNetflix}, 
we analyze a large data set of time-stamped video ratings from the video rental 
service Netflix that we model as a bipartite network of people and videos.  
In Section \ref{burst}, we examine the bursty behavior of individual users.  In 
Section \ref{SecCatalog}, we develop a catalog model of bipartite network growth and 
evolution. We then study the Netflix data using this model in 
Section \ref{SecNetflixCatalog}. Finally, we discuss our results and present 
directions for future research in Section \ref{SecConclusions}.


\section{Netflix Video Ratings} 
\label{SecNetflix}

\begin{figure}
  \includegraphics[width=250px]{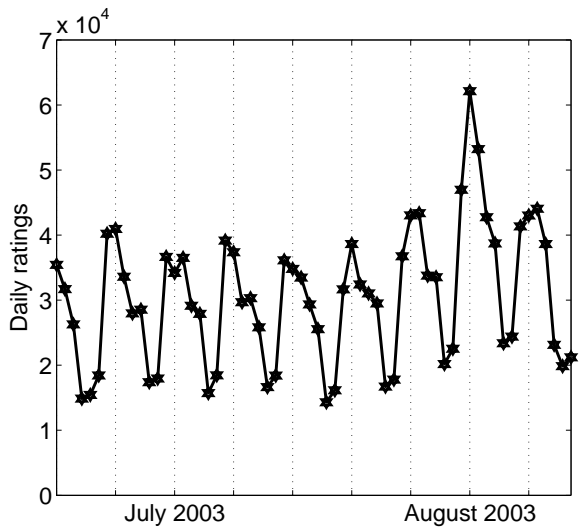}
  \caption{Number of daily ratings for each day in July and August 2003.  
  The mean number of ratings per day over this period is 30,449.  The dashed 
  vertical lines indicate Tuesdays.}
  \label{RatingsPerDayJulAug}
\end{figure}

Netflix is an online video rental service that encourages its users to rate the videos 
they rent in order to improve their personalized recommendations. As part of the 
{Netflix Prize} competition \cite{Netflixprize}, in which the company 
challenged the public to improve their video recommendation algorithm, Netflix 
released a large, anonymized collection of user-assigned ratings of videos in its 
catalog.  In this paper, we use the Netflix data to study human dynamics in the form 
of video ratings from a limited catalog.  One can also examine the dynamics of the 
ratings themselves, which would complement a recent empirical study of video ratings 
that used data from the Internet Movie Database (IMDB) \cite{imdb}.  The Netflix data 
consist of 100,480,507 ratings of 17,770 videos.  The ratings, which were given 
by 480,189 Netflix users between October 1998 and December 2005, were sampled 
uniformly at random by Netflix from the set of users who had rated at least 
20 videos \cite{netflixPaper}. 
Each entry in the data includes the video ID, user ID, rating score (an integer 
from 1 to 5), and submission date.  To illustrate some of the temporal dynamics in 
the data, we show in Fig.~\ref{RatingsPerDayJulAug} the total number of ratings for 
each day from July to August 2003.  The number of daily ratings exhibits a weekly 
pattern in which Mondays and Tuesdays have the highest activity and Saturdays and 
Sundays have the lowest. This reflects the weekly patterns in human work--leisure
habits.

\begin{figure}[htp]
  \begin{center}
    \includegraphics[width=250px]{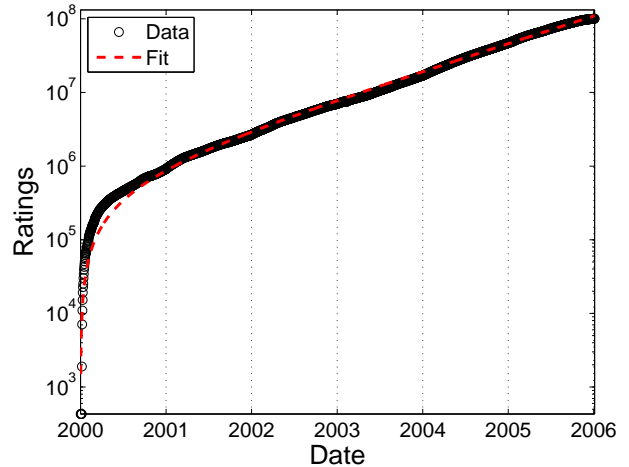}
    \caption[Growth of the ratings]{(Color online) Number of ratings in
      the Netflix data versus time from the beginning of 2000 to the end of 2005.
    Circles indicate data from Netflix and the dashed red curve is a fit to
    equation (\ref{RatingsGrowthTimeEq}).}
    \label{RatingsGrowthTime}
  \end{center}
\end{figure}

Figure~\ref{RatingsGrowthTime} shows the total number of ratings from 2000 to the 
end of 2005.  These ratings seems to grow exponentially, which we confirm by fitting 
the data to the function
\begin{equation}
  	r(t) = a_r\left(e^{b_r t} - 1\right)
  \label{RatingsGrowthTimeEq}
\end{equation}
using nonlinear least squares. We obtain the parameter values 
$a_r \approx 6.3656 \times 10^5$ and $b_r \approx 0.0024$. 

\begin{figure}[htp]
  \begin{center}
    \includegraphics[width=250px]{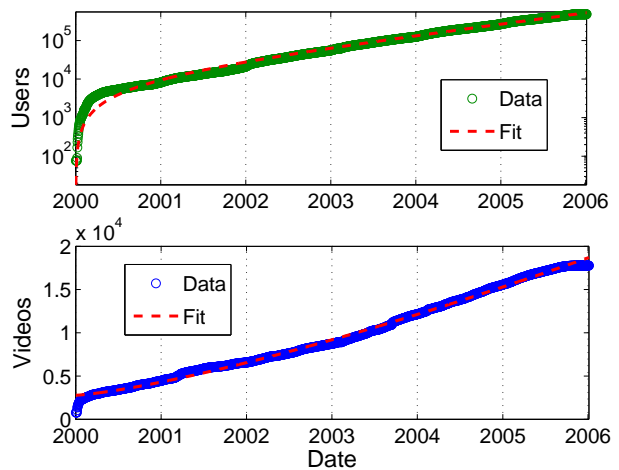}
    \caption[Growth of the users]{(Color online) Number of users (top) and videos
      (bottom) in the Netflix data versus time from the beginning of 2000 to the end 
      of 2005. Circles indicate data from Netflix and the dashed red curves are 
      fits to  equations (\ref{UsersGrowthTimeEq}) and (\ref{MoviesGrowthTimeEq}) for
    users and videos, respectively.}
    \label{UsersMoviesGrowthTime}
  \end{center}
\end{figure}

The number of users also grows exponentially, as shown on the top panel of 
Fig~\ref{UsersMoviesGrowthTime}.  The dashed curve in the plot is the fit to
\begin{equation}
  	u(t) = a_u(e^{b_ut} - 1)\,,
  \label{UsersGrowthTimeEq}
\end{equation}
where we obtain $a_u \approx 1.0018 \times 10^4$ and $b_u \approx 0.0018$.  We will 
need to take the exponential growth of the system into account when comparing data 
from dates that are far apart from each other. 

In the bottom panel of Fig.~\ref{UsersMoviesGrowthTime}, we show the number 
of videos from 2000 to 2005. The number of videos appears to grow roughly linearly 
as a function of time, but in fact it is better described by the relation
\begin{equation}
  	m(t) = a_m + b_mt^{c_m}\,,
  \label{MoviesGrowthTimeEq}
\end{equation}
where fitting yields $a_m \approx 2780.00$, $b_m \approx 0.6705$,
 and $c_m \approx 1.3097$.


\subsection{Bipartite Network Formulation}

The Netflix data can be represented as a bipartite network.  The two types of 
nodes in this network are users and videos.  We use $\mathcal{U}$ to denote the 
set of users and $\mathcal{M}$ to denote the set of videos.  
We ignore the rating values and consider only the presence or absence of a rating 
event, which corresponds to an edge between a user and a video in the unweighted  
bipartite network. The large size and longitudinal nature of the data provides 
a valuable opportunity to study video rating in the context of {human dynamics}, as 
has been done previously with mobile 
telephone networks \cite{J.P.Onnela05012007,nature06958}, book sale rankings 
\cite{Sornette.93.228701}, and electronic and postal mail usage patterns 
\cite{barabasi-2005-435, oliveira2005437}.


\subsection{Degree Distributions}

\begin{figure}
  \begin{center}
      \includegraphics[width=250px]{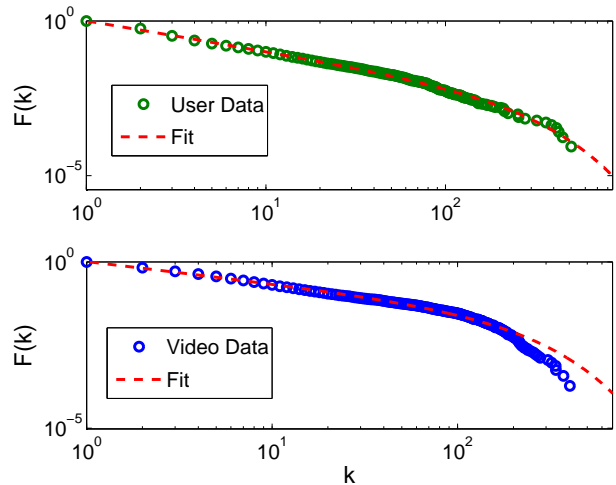}
  \end{center}
  \caption{(Color online) Cumulative degree distributions of user (top) and video 
  (bottom) nodes for August 26, 2003 (a Tuesday).  The dashed curves are the fits 
  to equation (\ref{PL-cutoff}) with parameters 
  $a \approx 0.9828$, $b \approx 0.0057$ for the users and 
  $a \approx 0.6622$, $b \approx 0.0070$ for the videos.}
  \label{DDweek34-2003}    
\end{figure}

The bipartite video-rating network has one degree distribution for the 
user nodes and another one for the video nodes.  Keeping in mind the observations in 
Fig.~\ref{RatingsPerDayJulAug}, we examine the cumulative degree distributions of 
individual days. The distributions have a similar functional form for 
each day in the data set.  We fit them to a power law with an exponential cutoff, 
\begin{equation}
  	F(k) \sim k^{-a}e^{-bk}\,,
  \label{PL-cutoff}
\end{equation}
using a modification of the method discussed by Clauset {\it et al.} 
\cite{Clauset:2007p5520}.  As an example, we show in Fig.~\ref{DDweek34-2003} the 
cumulative degree distributions for one day.  Table~\ref{CoeffsTable} gives the 
parameter values that we found in our fits of the data to equation (\ref{PL-cutoff}). 
Despite the weekly pattern of the ratings shown in Fig.~\ref{RatingsPerDayJulAug}, 
we did not find any significant differences between the values of $a$ and $b$ for 
different days of the week.  Hence, although the number of daily ratings does differ 
significantly among weekdays, such differences seem to not have much effect on the 
aggregate structure of the network.

The problem setting sheds some insight into the observed functional form of the degree 
distribution.  Users select which videos to rate from a large set of possibilities and 
possess time limitations on the number of videos that they are able to watch and rate. 
As in any market, videos must compete against each other for users' attention. 
One can also anticipate that certain videos saturate their market, 
especially in the case of niche videos whose audience is small to begin with.   
Once the demand for a niche video has been met, it virtually ceases to receive further 
ratings.  On the other hand, blockbusters might continue receiving numerous 
ratings for a long period of time.

\begin{table}
  \centerline{
    \begin{tabular}{|l|c|c||c|c|}
      \hline & \multicolumn{2}{c||}{$a$} & \multicolumn{2}{c|}{$b$} \\
      \hline & Mean & Var & Mean & Var \\
       \hline Videos & 0.6580 & 0.0200 & 0.0686 & 0.0100 \\
       \hline Users & 0.8381 & 0.0573 & 0.0116 & 0.0007 \\
      \hline 
  \end{tabular}}
  \caption{Fitting parameters of the daily video and user degree distributions 
    from 2000 to 2005 for the power law with exponential cutoff in 
    (\ref{PL-cutoff}).}
  \label{CoeffsTable}
\end{table}


\subsection{Clustering coefficients}

\begin{figure}
  \includegraphics[width=170px]{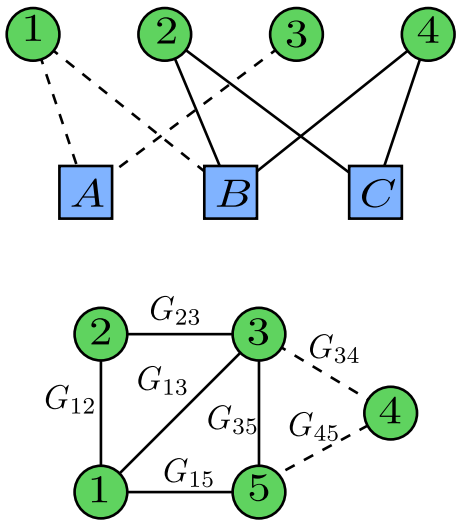}
  \caption{(Color online) Examples of how to calculate clustering coefficients 
  for bipartite (top) and unipartite (bottom) networks.  In the bipartite network, 
  solid lines indicate edges that form the square that includes node $B$, whose 
  bipartite clustering coefficient calculated according to equation (\ref{C4}) is 
  $C_4 = 1/5$.  One obtains this result because there are five possible squares for 
  this node (\{$1A2B$, $1C2B$, $1A4B$, $1C4B$, $2C4B$\}) but only one of them 
  ($2C4B$) actually exists. In the unipartite network, the solid lines indicate 
  edges that form the triangles that include node $1$.  If this were an unweighted 
  network, for which $G_{ij} \in \{0,1\}$ for all $i$ and $j$, then one would obtain an 
  unweighted clustering coefficient of $C_3(1) = 2/3$.  To calculate the value of 
  its weighted clustering coefficient $\tilde{C}_3$, we use equation (\ref{C3}).} 
   \label{ccSchema}
\end{figure}

To investigate the local connectivity of nodes and examine the impact
of highly-connected nodes, we calculate bipartite clustering coefficients 
\cite{martacluster,Zhang20086869}. In bipartite networks, a clustering coefficient for 
a node can be calculated by counting the number of cycles of length 4 (i.e., the 
number of ``squares'') that include the node and dividing the result by the total 
possible number of squares that could include the node. As stated by 
Zhang~\textit{et al.} in~\cite{Zhang20086869}, the possible (or underlying) number 
of squares is calculated 
by adding the potential links (including existing ones) between a particular node and
the neighbors of its neighbors. In Fig.~\ref{ccSchema} we show how
a square occurs in a bipartite network when two neighbors of 
a node have another neighbor in common. Bipartite networks cannot have triangles 
(three mutually-connected nodes) because two nodes of the same type cannot be 
neighbors, so a square is the shortest possible cycle.

The definition of a clustering coefficient of node 
$m_i$ in an unweighted bipartite network is~\cite{Zhang20086869}:
\begin{equation}
  	C_4(m_i) = \frac{\sum_{h,j}{q_{i_{jh}}}}
        {\sum_{j,h}{\left[(k_j - \eta_{i_{jh}})+(k_h- \eta_{i_{jh}}) + q_{i_{jh}}\right]}}\,, 
  \label{C4}
\end{equation}
where $q_{i_{jh}}$  is the observed number of squares containing $m_i$ and
any two neighbors $u_h$ and $u_j$. The degrees of the neighbors are $k_h$ and $k_j$, 
respectively, and $\eta_{i_{jh}}= q_{i_{jh}} + 1$. The possible number of squares is 
calculated adding the degrees of the nodes $u_h$ and $u_j$ minus the link that each 
shares with $m_j$ if the three nodes are not part of a square to avoid 
double-counting. If the three nodes are part of a square, then the square 
represented by the deleted link must be added again, hence 
$(k_j - \eta_{i_{jh}})+(k_h- \eta_{i_{jh}}) + q_{i_{jh}}$ in the denominator
of equation (\ref{C4}).

\begin{figure}
  \includegraphics[width=250px]{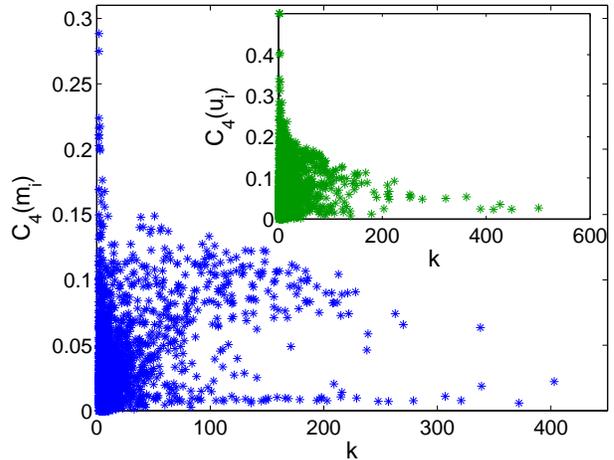}
  \caption{(Color online) Bipartite clustering coefficients $C_4(m_i)$ for video 
    (blue) and  user nodes (inset, green) for   August 12, 2003 (a Tuesday). 
    The mean values for this day are 
    $\langle C_4 \rangle = \frac{1}{M}\sum_{i=1}^{M}C_{4}(m_i) \approx 0.02606$ 
    for the videos and 
    $\langle C_4 \rangle = \frac{1}{U}\sum_{i=1}^{U}C_{4}(u_i) \approx 0.03144$ 
    for the users.}
  \label{C4Movies12Aug2003}
\end{figure}



In Fig.~\ref{C4Movies12Aug2003}, we show the values of $C_4(m_i)$ for the video 
and user nodes for a single day (Tuesday, August 12, 2003).  
In Table \ref{Table2003C4C3}, we show the mean values of the bipartite 
clustering coefficient of all one-day snapshots of Netflix in 2003. In spite of the
weekday-dependent variation in the number of daily ratings, the values of the 
bipartite clustering coefficient do not vary significantly across weekdays. 
However, the values of $\langle C_4 \rangle$ increase almost by 80\% for both 
node-types on weekends. For a network constructed from a single day's data, only 
about 2\% of the possible squares typically exist; this is comparable to what would 
occur in a random network with the same degree distributions. To investigate whether 
the presence of blockbuster nodes (which have high degrees and increase considerably 
the number of possible squares) has any effect on the value of  
$\langle C_4 \rangle$, we calculated the clustering coefficient after removing 
the top ten most rated videos. We did not find any conclusive evidence of 
blockbusters driving the value of the clustering coefficient; some of them caused 
the value of $\langle C_4 \rangle$ to go down and others caused it to go up.

One can also examine clustering coefficients in the weighted unipartite networks 
given by the projected adjacency matrices 
$\mathbf{G}_{\mathcal{U}}$ and $\mathbf{G}_{\mathcal{M}}$.  
We calculate the weighted clustering coefficient for each projection using 
the formula 
 \cite{PhysRevE.71.065103}
\begin{equation}
  	\tilde{C}_3(m_i) = \frac{2}{k_i(k_i-1)}\left[\frac{1}{G_M}\sum_{j,h} 
        \left(G_{ij}G_{ih}G_{hj}\right)^{1/3}\right]\,,
  \label{C3}
\end{equation}
where $k_i$ is again the degree of node $m_i$, $G_{ij}$ is the weight of 
the edge between $m_i$ and $m_j$, and $G_M = \max (G_{ij})$ denotes the maximum edge 
weight in the network.  The geometric mean $(G_{ij}G_{ih}G_{hj})^{1/3}$ of the edge 
weights give the ``intensity" of the $(i,j,h)$-triangle. When the network is 
unweighted, $(G_{ij}G_{ih}G_{hj})^{1/3}$ is $1$ if and only if all 
edges in the $(i,j,h)$-triangle exist and $0$ if they do not, reducing the equation 
to the unweighted unipartite clustering coefficient
\begin{equation}
    C_3(m_i) = \frac{2t_i}{k_i(k_i-1)}\,, \label{C3uw}
\end{equation}
where $t_i$ is the number of triangles that include node $m_i$.

\begin{figure} 
  \includegraphics[width=250px]{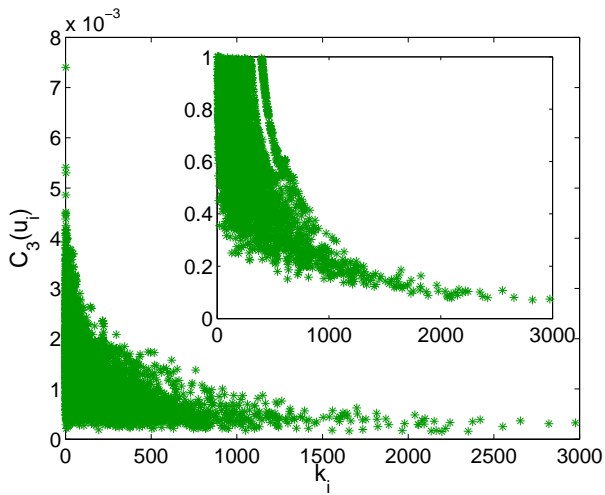}
  \caption{Weighted clustering coefficient $\tilde{C}_3(u_i)$ for nodes in the 
  unipartite projection onto users for August 4, 2003. The $x$-axis represents 
  node degrees, and the $y$-axis represents $\tilde{C}_3(u_i)$.  The mean values 
  for this day are  
  $\langle \tilde{C}_3 \rangle =\frac{1}{U}\sum_{i=1}^{U} \tilde{C}_{3}(u_i) 
  \approx 0.0013$ for the projection onto users and $\tilde{C}_3 \approx 0.0086$ for 
  the projection onto videos (not shown). The inset shows values of the unweighted  
  coefficient $C_{3}(u_i)$ from the same data.}
  \label{C3-04Aug2003}
\end{figure}

\begin{table}
  \centerline{
    \begin{tabular}{|l|c|c||c|c|} \hline
      & \multicolumn{2}{c||}{$\langle C_4\rangle$} 
      & \multicolumn{2}{c|}{$\langle \tilde{C}_3 \rangle$}  \\
      \hline & mean & var & mean & var \\
      \hline Videos & 0.02039 & 0.0007 & 0.0056 & $10^{-6}$ \\
      \hline Users & 0.02092 & 0.0012 & 0.0044 & $10^{-6}$ \\
      \hline
  \end{tabular}}
  \caption{Means and variances of $\langle C_4 \rangle$ (for the bipartite network) 
    and $\langle \tilde{C}3 \rangle$ (for the projections)
    of videos and users on single-day snapshots of 2003, calculated using 
    equations~(\ref{C4}) and~(\ref{C3}).}
  \label{Table2003C4C3}
\end{table}

In Fig.~\ref{C3-04Aug2003}, we show the $\tilde{C}_3(u_i)$ values for the user 
projection $\mathbf{G}_{\mathcal{U}}$ (with 10,228 nodes and 814,667 edges) from 
Tuesday, August 4, 2003.
In Table \ref{Table2003C4C3}, we show the mean clustering-coefficient 
values for the projected user and video networks for all single-day snapshots of 2003.
The values of $\langle \tilde{C}3 \rangle$ did not vary much among weekdays, except
for the videos' $\langle \tilde{C}3 \rangle$ that almost doubled its value on the 
weekends from an average of 0.0045 from Monday to Friday to 0.0086 on Saturday and 
Sunday.


Given the values of $\langle C_4 \rangle$ in Table \ref{Table2003C4C3}, it is 
unsurprising that the values of $\langle \tilde{C}_3 \rangle$ are also typically low.
In the inset of Fig.~\ref{C3-04Aug2003}, we show the values of the users' 
unweighted clustering coefficient $C_3$, which are naturally much higher.  For 
example, about 4000 users have $C_3 = 1.0$, indicating that all potential triangles 
exist among these users.  This differentiates one set of nodes from the rest. 
This feature, which we observe often in the data, arises from the dominant video 
of the day.  For August 4, 2003, this video (which is typically a blockbuster) 
was Daredevil, which had 396 ratings and created many edges in the user projection 
among the users who rated it. Removing Daredevil from the bipartite network also 
removes these deviant nodes.  This feature is not apparent if one calculates only 
the unweighted unipartite clustering coefficient ${C}_3$. 
Just as we did with $\langle C_4 \rangle$, and given the dramatic effect observed 
by removing Daredevil, we calculated $\langle C_4 \rangle$ for the projected network 
of users removing the ten most rated videos. We found that for every additional 
video removed, the value of $\langle C_3 \rangle$ increased by 0.2\%, while for 
$\langle \tilde{C}_3 \rangle$ the increment was slightly larger.


\section{User Bursts}\label{burst}

\begin{figure}
  \includegraphics[width=250px]{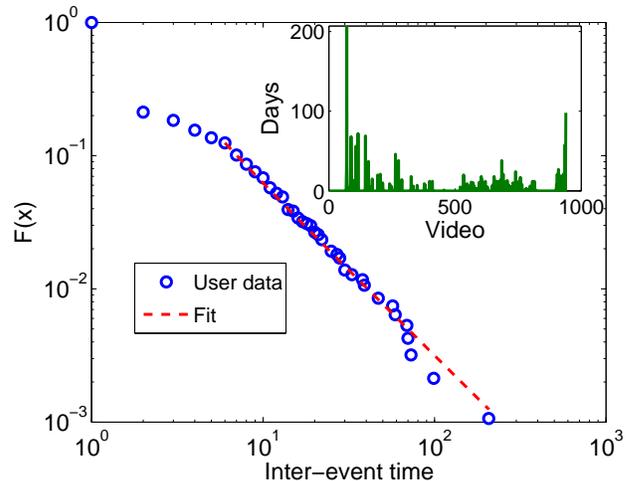}
  \caption{(Color online) Cumulative distribution of the inter-event time between
    the ratings of one Netflix user. The user signed up on April 4, 2000, and 
has a degree of 940 based on ratings cast over a period of almost five years.
  The dashed curve indicates the fit to the function $F(x) \sim x^{-\alpha}$, which 
  yields $\alpha \approx 2.27$ in this case. The inset shows the number of days
  between consecutive video ratings.}
  \label{User46DelayDist}
\end{figure}

A close examination of the rating habits of individual users can also yield rich and 
informative insights.  Recent research has shown that people tend to have bursts of 
e-mail and postal correspondence, in which they send and receive numerous messages 
within short periods of time, followed by long periods of inactivity 
\cite{barabasi-2005-435,oliveira2005437,burstbook}. We find similar features in 
the Netflix data, as about 70\% of the users exhibit bursty behavior by rating 
several videos in one go after several days of no activity. We illustrate this 
phenomenon in Fig.~\ref{User46DelayDist} by plotting the cumulative distribution of 
inter-event times between the ratings of one user over a period of almost five years. 
We fit this distribution to a power law 
$F(x) \sim x^{-\alpha}$ using the method discussed in Ref.~\cite{Clauset:2007p5520} 
to determine the value of the exponent $\alpha$.  We can similarly provide estimates 
for possible power laws (with actual power laws over roughly two decades of data) 
among the other bursty users, though the value of $\alpha$ depends on the final degree 
(i.e., the total number of rated videos) of the user.  
For example, bursty users with final degrees between 100 and 1000 have a mean 
exponent of $\alpha \approx 2.54$, whereas those with final degrees of 
at least 4000 have a mean exponent of $\alpha \approx 3.17$.  
Additionally, there are several types of users among those who do not exhibit bursty 
dynamics.  In particular, some users rated only a very small number of videos (which 
may be due to the sampling done by Netflix) and others exhibit seemingly unrealistic 
levels of rating activity. (For example, there are 47 users who signed up in January 
2004 or later and who have rated more than 4000 videos each.) 


\section{Catalog Networks} \label{SecCatalog}

The above empirical investigation of the Netflix data motivates the development of an 
evolution model for bipartite {\it catalog networks}, which arise in a diverse set
of applications. Such networks have two sets of nodes whose numbers can be fixed or 
dynamic, and edges are placed one at a time between previously unconnected
edges that are chosen according to predefined rules. One continues to add edges until
a predefined final time has been reached or the system has become saturated, at
which point every node in one partite set is connected to every node in
the other partite set. The Netflix network can be studied using such a catalog network
framework; it starts completely disconnected (nobody has rated any videos), and the 
users start choosing and rating videos from the catalog. Depending on the way 
the data set is sampled, the catalogs can be static (e.g. a one-day snapshot) or
dynamic (e.g. the full data set).
Catalog models of network evolution are closely related to the 
network rewiring problem studied by Plato and Evans \cite{evans:056101,evans-2008-3} 
that features fixed sets of artifacts and individuals. Every individual has one 
affiliation (a connection) with an artifact and can reassign this connection to 
another node as the network evolves. In contrast, in a catalog network, any edge that 
has been placed between two nodes in the network is permanent.  Consequently, catalog 
networks are suited to describing records of interactions that are assigned 
dynamically and then remain permanently in the system.

As before, $\mathcal{U}$ denotes the set of users and $\mathcal{M}$ denotes the set 
of videos. The size of $\mathcal{U}$ is $u(r)$ and the size of $\mathcal{M}$ 
is $m(r)$, where $r$ denotes a discrete time that is indexed by the ratings.  
That is, we take every rating event as a time step, so when we discuss time 
in this context, we are referring to ``rating time'' and not physical time unless 
we indicate otherwise. 
Because $m(r)$ and $u(r)$ are not always integers, we define $U(r) = \floor{u(r)}$ 
and $M(r) = \floor{m(r)}$ as the (nonnegative integer) numbers of user and video 
nodes, respectively.
The associated time-dependent catalog vectors, $D_{\mathcal{U}}$ and $D_{\mathcal{M}}$, have 
components given by the degrees of each node in the catalog:
\begin{equation}
  	D_{\mathcal{U}}(r) =
  \begin{bmatrix}
    	k_{u_1}(r) \\
    	k_{u_2}(r) \\
    	\vdots \\
    	k_{u_U(r)}(r)
  \end{bmatrix}, \qquad
  D_{\mathcal{M}}(r) =
  \begin{bmatrix}
    	k_{m_1}(r) \\
    	k_{m_2}(r)\\
    	\vdots \\
    	k_{m_M(r)}(r)
  \end{bmatrix}\,.  \label{degVecs}
\end{equation}
These vectors have size $U(r)$ and $M(r)$, respectively.
We denote by $N_{\mathcal{U}}(r,k)$ (with $k \in \{0,1, \hdots, M(r)\}$) 
and $N_{\mathcal{M}}(r,k)$  (with $k \in \{0, 1, \hdots, U(r)\}$) the numbers of 
users and videos, respectively, that have degree $k$ at rating time $r$.  One can
normalize $N_{\mathcal{U}}(r,k)$ to obtain the proportion of nodes with degree 
$k$ given by $\hat{N}_{\mathcal{U}}(r,k) = \frac{1}{U(r)}N_{\mathcal{U}}(r,k)$. 
An analogous relation holds for $\hat{N}_{\mathcal{M}}(r,k)$. 

Based on our intuition about the choosing and rating of videos, we add edges to the 
network using a combination of linear preferential attachment and uniform attachment.
On one hand, one expects the choice of a user to be driven in part by the choices 
made by others, as popular videos are more likely to attract further viewings and 
hence ratings. On the other hand, one also expects an element of idiosyncrasy on 
the part of each user, allowing him or her to choose any video from the catalog 
regardless of the choices of others. This results in two time-dependent 
probabilities---one for users and one for videos---each of which consists of a 
convex combination of preferential and uniform attachment. More specifically, each 
time an edge is added 
to the network, we select a user and a video to be connected by this new edge. The 
video (user) node is chosen using uniform attachment with probability $1-q$ 
(respectively, $1-p$) and linear preferential attachment with probability $q$ 
(respectively, $p$). The addition of an edge occurs during a single discrete 
(rating) time step, as is common in models of network evolution. Combining 
these ideas, a video node with degree $k_i$ is chosen with probability
\begin{align}
  	P_{\mathcal{M}}(r, k_i) & = \frac{1-q}{M(r) - N_{\mathcal{M}}(r, U(r))} \nonumber \\
        & \quad +  \frac{q k_i}{\|D_{\mathcal{M}}(r)\|_1 - U(r) N_{\mathcal{M}}(r, U(r))}\,,
    \label{pqProbMovie} 
\end{align}
and a user node with degree $h_i$ is chosen with probability
\begin{align}
  	P_{\mathcal{U}}(r, h_i) & = \frac{1-p}{U(r) - N_{\mathcal{U}}(r, M(r))}  \nonumber \\
        & \quad + \frac{p h_i}{\|D_{\mathcal{U}}(r)\|_1 - M(r) N_{\mathcal{U}}(r, M(r))}\,, 
      \label{pqProbUser}
\end{align}
where the values of the parameters $p, q \in [0, 1]$ are fixed, 
$\|D_{\mathcal{U}}(r)\|_1 = \sum_{i=1}^{U(r)}k_i(r)$, and 
$\|D_{\mathcal{M}}(r)\|_1 = \sum_{i=1}^{M(r)}h_i(r)$.  The probabilities 
$P_{\mathcal{U}}(r, h_i)$ and $P_{\mathcal{M}}(r, k_i)$
change over time as the degrees of the nodes change when edges are added to 
the network.

The denominators in equations (\ref{pqProbMovie}-\ref{pqProbUser}) contain the terms 
$N_{\mathcal{M}}(r, U(r))$ and $N_{\mathcal{U}}(r, M(r))$ 
because once a node of either type is fully connected, it is no longer eligible to 
receive any new connections and is effectively no longer in the catalog until a new
node of the other type arrives. 
When $r = 0$, one obtains $\|D_m(0)\|_1 = \|D_u(0)\|_1 = 0$ and 
$N_{\mathcal{M}}(0, U(r)) = N_{\mathcal{U}}(0, M(r))=0$, which would result in 
division by zero.  To overcome this problem, we 
follow the standard procedure employed in network growth models \cite{albert-2002-74} 
by seeding the algorithm with an edge that connects two randomly-chosen nodes (one 
from each of the partite sets).  This is equivalent to shifting the rating-time 
variable and changing the initial conditions to $\|D_m(0)\|_1 = \|D_u(0)\|_1 = 1$. 


\subsection{Rate Equations}

One can use rate equations (i.e., master equations) to investigate the dynamics of the 
degree distributions of a catalog network.  This type of approach has been used 
successfully to study a variety of other networks 
\cite{PhysRevLett.85.4629, evans:056101, evans-2008-3,       
PhysRevLett.86.3200, PhysRevE.71.036127, Newman:2003}. 
The analysis of the degree distribution for videos in the catalog model is identical 
to the one for users, as only the constants and sizes of the catalogs are different.  
Accordingly, we present our results for the degree distributions of the videos; one 
obtains the results for user distributions by changing $q$ to $p$, $M(r)$ to 
$U(r)$, and $P_{\mathcal{M}}(r, k)$ to $P_{\mathcal{U}}(r, k)$. For notational convenience, 
we also drop the subscripts in this subsection,  
so $N(r, k)$ denotes the number of nodes with degree $k$ at time $r$. To construct 
the rate equations, one must consider how many nodes pass through $N(r, k)$ (i.e.
turn into nodes of degree $k$ and $k+1$) for $k \in \{0,1,2,\hdots, U(r)\}$. 
This yields 
\begin{align}
  \ddr{N(r, 0)} &= m'(r) -P_{\mathcal{M}}(r, 0)N(r, 0)\,, 
  \nonumber \\
  \ddr{N(r, k)} &= P_{\mathcal{M}}(r, k-1)N(r, k-1) 
   \label{pqNkMasterEqGrwCat} \\
  &- P_{\mathcal{M}}(r, k)N(r, k)\,, 
  \quad k > 0,\nonumber
\end{align}
where $m'(r) = \ddr{m(r)}$. The initial conditions are
\begin{align}
  N(0, 0) &= M(0)-1\,, \nonumber \\
  N(0, 1) &= 1\,, \label{initCondGrwCat} \\
  N(0, k) &= 0\,, \quad k >1\,. \nonumber
\end{align}
Equation (\ref{pqNkMasterEqGrwCat}) is a system of coupled nonlinear 
ordinary differential equations (ODEs). The positive and negative terms account,
respectively, for an increase and decrease in the number of nodes of a given 
degree as nodes receive new edges.  The equation 
for $N(r, 0)$ has $m'(r)$ as a positive term to indicate the entry of new nodes 
(with degree $0$) to the network. The time-dependent probabilities 
$P_{\mathcal{M}}(r, k)$ are defined in equation (\ref{pqProbMovie}).  In the case of 
fixed catalogs, there is a maximum value of $k$, so the final equation in 
(\ref{pqNkMasterEqGrwCat}) takes a slightly different form (see below).


\subsubsection{Fixed Catalogs}

We begin by analyzing the evolution of the network with fixed catalog sizes, 
so $U(r)=U$, $M(r)=M$, and $m'(r)=0$ for all $r$. Because a 
finite, fixed number of users and videos are available in the catalogs, the 
network can only evolve until time $r=UM$.  At this point, the system becomes 
saturated (i.e., $N_{\mathcal{U}}(MU, M) = U$ and $N_{\mathcal{M}}(MU, U) = M$), 
and no additional edges can be added to the network.
Note additionally that the equations in (\ref{pqNkMasterEqGrwCat}) change slightly for 
fixed catalogs. In particular, the last equation for nodes with degree
$U$ changes to
\begin{equation}
   \ddr{N(r, U)} = P_{\mathcal{M}}(r, U-1)N(r, U-1)\,,
   \label{pqNkMasterEqDegU}
\end{equation}
which only has one positive term because nodes with degree $U$ stay that way until the
end of the process.

Additionally, while the degree distribution of a network generated using the catalog 
model with static node sets is time-dependent, the long-time asymptotic behavior 
is always the same:
\begin{equation}
  \lim_{r \to UM}N(r, k) =
  \left\{
  \begin{array}{lll} 
    M\,, & \mathrm{if} & k=U\,, \\
    0\,, & \mathrm{if} & k < U\,,
  \end{array}
  \right.
  \nonumber
\end{equation}
which gives a de facto final condition to the system in 
(\ref{pqNkMasterEqGrwCat}-\ref{pqNkMasterEqDegU}).
Accordingly, we examine degree distributions for $r \leq UM - 1$.

\begin{figure}
      \includegraphics[width=250px]{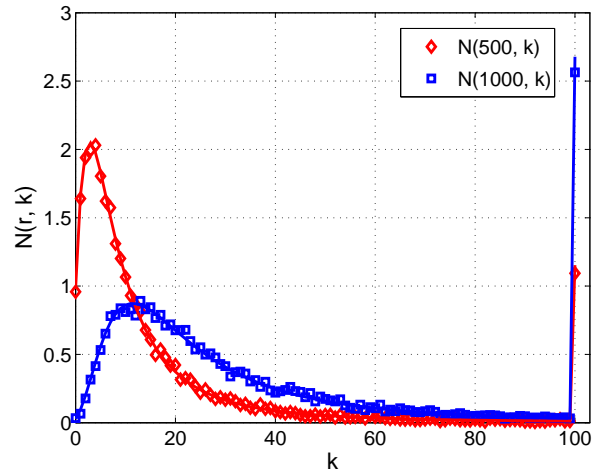}
      \caption{(Color online) Degree distributions of video nodes averaged over 500 
        simulations of a fixed catalog network with $U=100$ users, $M=30$ videos, and 
        $q=0.8$ at rating times $r=500$ (red diamonds) and $r=1000$ (blue squares).  
        The solid curves are the solutions to the differential equation 
        (\ref{pqNkMasterEqGrwCat}).}  
      \label{pkt}
\end{figure}

\begin{figure}
  \includegraphics[width=250px]{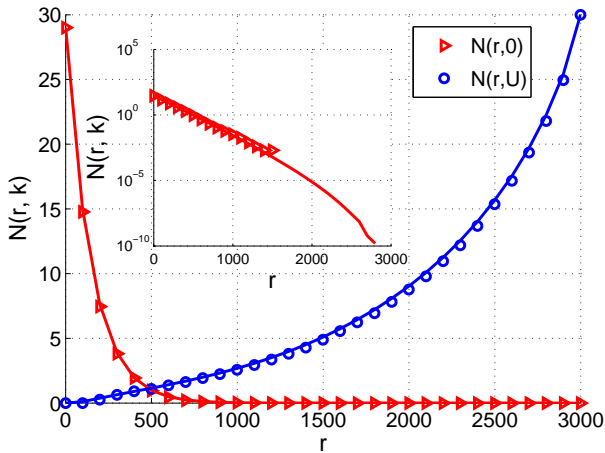}
  \caption{(Color online) Numbers of nodes $N(r, 0)$ with degree $0$ (red triangles) 
    and $N(r, 100)$ (blue circles)
    with degree $100$ from 500 simulations of a fixed catalog network with $U=100$, 
    $M=30$, and $p=0.8$. Inset: Decrease of $N(t, 0)$ on a 
    semi-logarithmic scale, which appears to decrease exponentially. 
    The solid curves come  from the solutions of (\ref{pqNkMasterEqGrwCat}).}
  \label{p0p100}
\end{figure}

In Fig.~\ref{pkt}, we show the degree distribution of the video nodes 
averaged over 500 simulations of a fixed catalog network with $U=100$ and $M=30$ at 
different times during its evolution.  As the discrete time $r$ increases, the 
peaks of the functions travels towards higher values of $k$ and decrease 
as if they were diffusing.  We also observe a jump in $N(r,k)$ at $k=U$.
This occurs because there are nodes in the network that become 
fully connected during the edge-assignment process (see Fig.~\ref{p0p100}). 
Interestingly, Johnson {\it et al.} showed recently that the 
time-dependent degree distributions observed in some 
networks that undergo edge rewiring with preferential attachment 
follow nonlinear diffusion processes \cite{PhysRevE.79.050104}. 

\begin{figure}
  \includegraphics[width=250px]{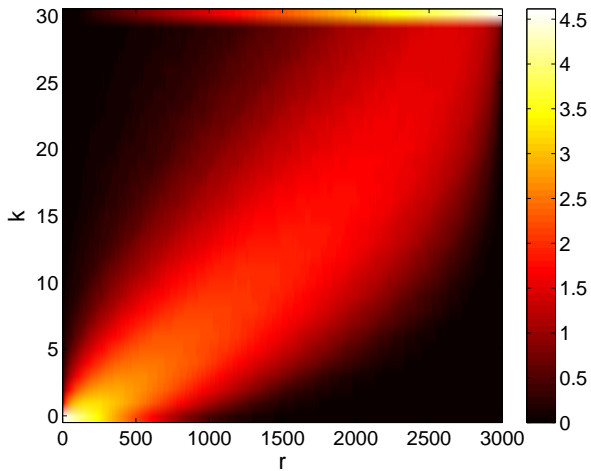}
  \caption{(Color online) Mean of $N(r, k)$ 
    for user nodes in 500 simulations 
    of a fixed catalog network with $U=100$, $M=30$, and $p=0.5$. 
    The axes are (rating) time $r$ and degree $k$, and the color indicates the 
    value of $\log(N(r, k)+1)$. 
    The horizontal line at the top of the image is the discontinuity (as seen
    with the video nodes in Fig.~\ref{p0p100}) that
    corresponds to the value of $N(r, M)$ and reflects the appearance of 
    fully-connected user nodes.}
  \label{resultsP05}
\end{figure}

Figure~\ref{resultsP05} reveals how the user nodes achieve full connectivity between 
$r = 0$ and $r = UM-1$.
The image  shows the ``paths'' that user nodes follow in the $(r,k)$-plane between 
$(0,0)$ and $(UM-1,M)$.  For example, the nodes that follow a 
steep (high $k$ for early $r$) trajectory are the ones that receive many links early 
on. Their degree grows mostly from preferential attachment in the edge-assignment 
mechanism, and they accordingly achieve full connectivity early in the process. 
The nodes that acquire edges more slowly initially begin to receive edges very fast as 
$r$ approaches $UM$ (because other nodes have already saturated), explaining the steep 
climb in the upper right corner of the figure.

\begin{figure}
  \includegraphics[width=250px]{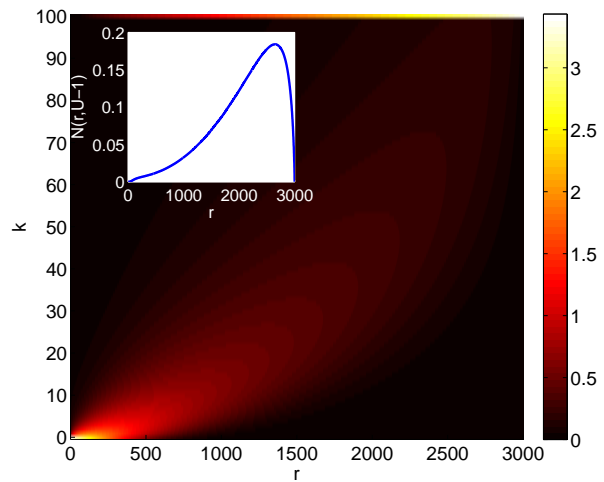}
  \caption{(Color online) Numerical solution of $N(r, k)$ for video nodes
    from equation (\ref{pqNkMasterEqGrwCat})  
    with a fixed catalog and $q=0.8$, $M=30$, and $U=100$.  
    (We again plot $\log(N(r,k)+1)$.) 
    The horizontal line at $k=100$ corresponds to the saturated nodes 
    $N(r, U)$. The inset shows a plot of $N(r,U-1)$ 
    for the same network.}
  \label{imagesPQCatUpdate}
\end{figure}

The ``final'' condition that $N(UM-1, U) = M$ makes the system in 
(\ref{pqNkMasterEqGrwCat}) very stiff for high values of $k$ and $r$.
Fig.~\ref{imagesPQCatUpdate} shows the path that the video nodes follow in the 
$(r,k)$-plane (i.e., the same information as in Fig.~\ref{resultsP05} but for 
video nodes) but for the numerical solutions of (\ref{pqNkMasterEqGrwCat}) instead of 
direct network simulations. In the inset of the Fig.~, we show the profile of 
$N(r, U-1)$ which evinces the aforementioned stiffness. Because all nodes must be 
fully connected at $r=UM-1$, nodes with low degrees begin to receive many edges for 
high values of $r$. This causes $N(r,k)$ for high $k$ to peak late in the process, 
and the nodes ``travel'' through values of $k$ rather quickly, which explains the 
incredibly steep slope of $N(r, U-1)$ as $r$ approaches $UM-1$.

The value of $q$ affects the width of the region (light colored) in the $(r,k)$ 
plane. For lower values of $q$ (e.g., $q=0.3$), uniform random 
attachment dominates and the region of activity becomes narrower.
The nodes attain edges at roughly the same pace. For larger values of $q$, the first 
nodes to receive edges become more likely to continue receiving more nodes until they 
saturate, 
and the area of activity of the nodes becomes wider 
(see Fig.~\ref{imagesPQCatUpdate}).


\subsubsection{Growing Catalogs}

In the previous section, we described the dynamics of catalog networks when
the sizes of the catalogs are fixed.  While this provides a good baseline 
investigation, catalogs can grow in many applications---for example, Netflix gains 
both new subscribers  and new videos almost every day. Accordingly, in this section 
we study the dynamics of (\ref{pqNkMasterEqGrwCat}) for growing catalogs for which 
$m'(r) > 0$.

\begin{figure}
\includegraphics[width=250px]{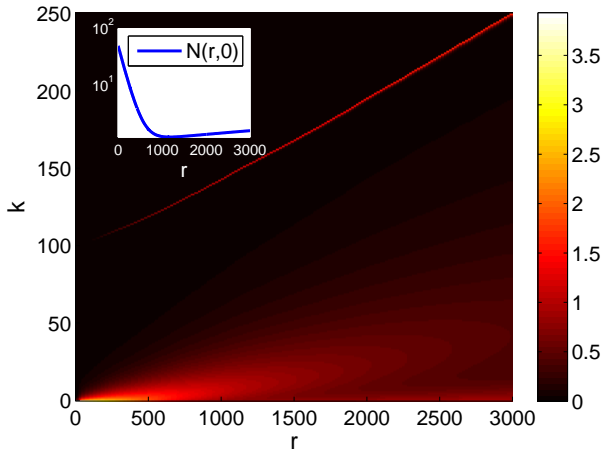}
\caption{(Color online) Numerical solution of $N(r, k)$ for video nodes
from equation (\ref{pqNkMasterEqGrwCat}) with $q=0.8$, $m(r)=30+ 0.007r$, 
and $U=100 + 0.05r$.  (We again plot $\log(N(r,k)+1)$.)  The increasing diagonal 
line gives $U(r)$ and represents the temporarily saturated nodes.  In the inset, 
we show a plot of $N(r,0)$ on a semilogarithmic scale. We observe a rapid initial 
decrease followed by a slower increase as the catalog grows.}
\label{pqP08GrwCat}
\end{figure}

The system no longer has an obligatory final time, and the saturation level of 
nodes is now time-dependent. For example, a user that has degree $M(r)$ 
is saturated temporarily until a new video ``arrives''---i.e., until time 
$r + \Delta r$ so that $M(r + \Delta r) - M(r) > 0$ and there is a new video to rate.

In Fig.~\ref{pqP08GrwCat}, we show a numerical solution to equation 
(\ref{pqNkMasterEqGrwCat}) where $m(r)$ and $u(r)$ are linear functions of $r$.
Instead of the horizontal line of fully connected nodes along $k=100$ in
Fig.~\ref{imagesPQCatUpdate}, the saturation of the nodes follows the
growth of $U(r)$. In the inset of Fig.~\ref{pqP08GrwCat}, we show the time profile of
$N(r,0)$. Initially, it has what appears to be exponential descent before 
it starts to grow slowly as the catalog size increases, in contrast to what we 
observed in Fig.~\ref{p0p100}. The early rapid decay is explained by the absence 
of many nodes with high degrees, so nodes with lower degrees receive edges.
As $r$ increases, the better-connected nodes receive more edges (because for 
$q=0.8$ the dominant mechanism is linear preferential attachment) 
and the population of nodes with fewer edges increases slowly. 
In Section \ref{SecNetflixCatalog}, we discuss how the Netflix data displays 
some of these features. 


\section{Netflix as a Catalog Network} \label{SecNetflixCatalog}

We now investigate how well our catalog model captures the 
human dynamics revealed by the Netflix data.  To do this, we sample the data 
set while keeping in mind the following considerations:
\begin{itemize}
\item Because of the way we have defined our catalog network growth
  model, we must consider the evolution of the Netflix data in ``rating time'', 
in which every new rating (which adds an edge in the network) constitutes a time step.
\item Although there might be a (physical) time difference between a node (either 
user or video) joining Netflix and the node receiving its first edge, this 
information is not included in the data. Many videos receive more than one rating 
on their first day, so their entry to the network is reflected by increases in 
the value of $N(r,k)$ for several values of $k$. We will have to take this into 
account when comparing our model to the data.
\end{itemize}


\subsection{Growth and Dynamics}

To compare our results to the data, we express the growth of the numbers of videos and 
users as a function of rating time $r$.  Solving for $t$ in equation 
(\ref{RatingsGrowthTimeEq}) gives
\begin{equation}
   t = \frac{1}{b_r}\log{\left(\frac{r}{a_r} + 1\right)}\,.
  \label{RatingTime}
\end{equation}
We substitute (\ref{RatingTime}) into (\ref{UsersGrowthTimeEq}) to obtain the 
new expressionfor the users as a function of ratings:
\begin{equation}
  	u(r) = a_u\left[\left(\frac{r}{a_r} +1\right)^{b_u / b_r} - 1\right]\,.
  \label{UsersRatingTimeEq}
\end{equation}
We follow the same procedure for the videos to obtain
\begin{equation}
  m(r) = a_m + b_m\left\{\frac{1}{b_r}\log{\left(\frac{r}{a_r} + 
    1\right)}\right\}^{c_m}\,.
  \label{MoviesRatingTimeEq}
\end{equation}

\begin{figure}[htp]
  \begin{center}
    \includegraphics[width=250px]{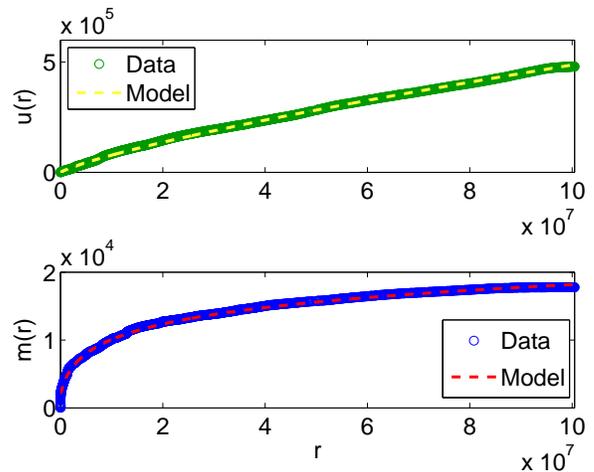}
    \caption[Growth of the users in rating-time]{(Color online) Users (top) and videos
      (bottom) as a function of ratings. We use circles to show the data from 
      Netflix and dashed curves to show the predictions from equations 
      (\ref{UsersRatingTimeEq}) and (\ref{MoviesRatingTimeEq}).  
      We use the parameter values obtained in Sec.~\ref{SecNetflix}. }
    \label{numUsersMoviesRatings}
  \end{center}
\end{figure}

In Fig.~\ref{numUsersMoviesRatings}, we show the numbers of users and videos
versus the number of ratings in the network.  Observe that the predictions
from equations (\ref{UsersRatingTimeEq}-\ref{MoviesRatingTimeEq}) agree
very well with the data.

\begin{figure}[htp]
  \begin{center}
    \includegraphics[width=250px]{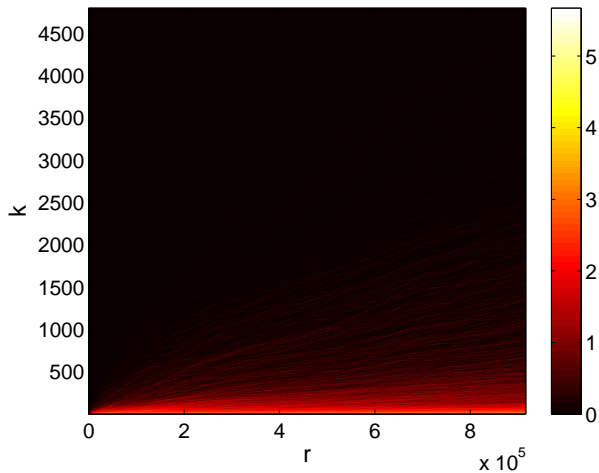}
    \caption{(Color online) Video degree distribution 
      $N_{\mathrm{data}}(r,k)$ in the Netflix data set in 2000.  (We again
      plot $\log(N_{\mathrm{data}}(r,k)+1)$.)  We show data for videos with degrees 
      ranging from 1 to 4794.}
    \label{MoviesYr2000AllDegs}
  \end{center}
\end{figure}

Figure~\ref{MoviesYr2000AllDegs} shows the time-dependent degree distribution
of videos in the Netflix data set for the year 2000. 
The sample in the plot consists of 365 measurements (one for each day) of
$r$ and $N(r, k)$. The highest degree in this sample is 4794; this is well below the 
theoretical maximum of 9289 according to the expression for $u(r)$ in equation 
(\ref{UsersRatingTimeEq}), so the network is not experiencing node saturation. 
We can rewrite the probability that a video node 
receives an edge as 
\begin{equation}
  P_{\mathcal{M}}(r, k_i) = \frac{1-q}{M(r)}  
  + \frac{qk_i}{\|D_{\mathcal{M}}(r)\|_1}\,. 
  \nonumber
\end{equation}
The rate equation for the evolution of the degree distribution is
\begin{align}
  	\ddr{N(r,1)}  = & \delta_1 m'(r) -  P_{\mathcal{M}}(r, 1)N(r,1)\,, \nonumber \\
  \ddr{N(r, k)} = & \delta_k m'(r) + P_{\mathcal{M}}(r, k-1)N(r,k-1) 
  \label{pqNkMasterEqGrwCatNonSat} \\
  &\quad - P_{\mathcal{M}}(r, k)N(r,k), \quad k > 1. \nonumber
\end{align}
The initial conditions are $N(0,1) = m(0)$ and $N(0, k) = 0$ for $k>1$.
As noted earlier, the lowest degree a node can have in the data is $1$ and the entry 
degree of the nodes can have any value of $k$. We denote by $\delta_k$ the proportion
of new nodes whose entry degree is $k$, such that $\sum_k \delta_k =1$. 
We investigated how many ratings do videos receive on the
day that they entered the system and found that over $97\%$ of the new nodes receive
$3$ or fewer ratings. Consequently, we have set $\delta_1=0.8$, $\delta_2=0.15$, 
and $\delta_3=0.05$.

To see how well our model describes the Netflix video data in the year 2000, we define 
$N_k(q)$ as the $4794 \times 365$  matrix obtained solving the system 
in (\ref{pqNkMasterEqGrwCatNonSat}) and $N_{\mathrm{data}}$  obtained from the data 
sample. These two matrices contain the values of $N(r, k)$ from the sample and 
from the equations for all values of $k$ and $r$. The matrices are of the given 
size because we sample the degree distribution once per day and the maximum degree 
observed is 4794. We define the error function
\begin{equation}
  	E(q) = \|N_k(q) - N_{\mathrm{data}}\|\,,
  \label{errorFunction}
\end{equation}
where $\| \cdot \|$ is the Euclidean matrix norm. To find the optimum value
$q^*$, we minimize $E(q)$
using the Nelder-Mead derivative-free simplex method~\cite{lagarias:112}. 
We found that the value of $q$ that minimizes (\ref{errorFunction}) 
is $q^* \approx 0.9795$, meaning that according to the model about $98\%$ of the 
decisions to rate a video by users are guided by its popularity (i.e., preferential 
attachment).

\begin{figure}[htp]
  \begin{center}
    \includegraphics[width=250px]{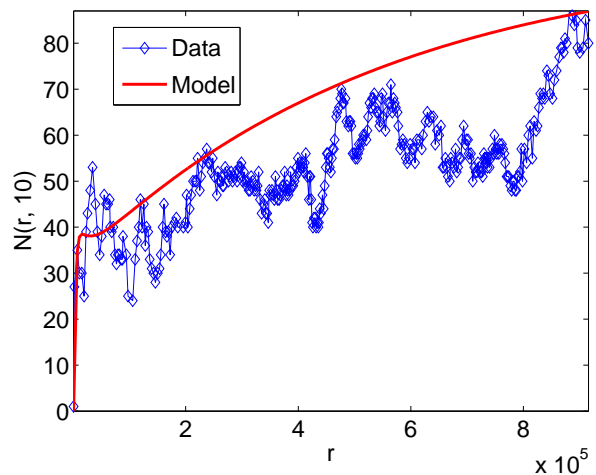}
    \caption{(Color online) Values of $N(r,10)$ (videos with degree 10) obtained 
      by solving (\ref{pqNkMasterEqGrwCatNonSat}) using $q = 0.9795$ (red curve) 
      and the data from Netflix that we report in Fig.~\ref{MoviesYr2000AllDegs} 
      (blue dots).} 
    \label{year2000deg10}
  \end{center}
\end{figure}

\begin{figure}[htp]
  \begin{center}
    \includegraphics[width=250px]{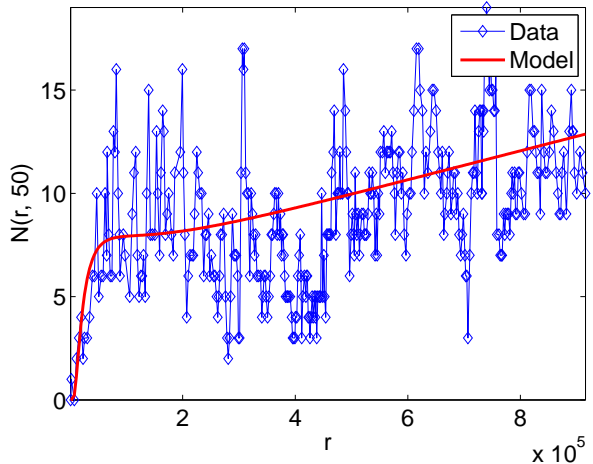}
    \caption{(Color online) Values of $N(r,50)$ (videos with degree 50) obtained by 
      solving (\ref{pqNkMasterEqGrwCatNonSat}) using $q = 0.9795$ (red curve) 
      and the data from Netflix that we report in Fig.~\ref{MoviesYr2000AllDegs} 
      (blue dots).} 
    \label{year2000deg50}
  \end{center}
\end{figure}

In Figs.~\ref{year2000deg10} and \ref{year2000deg50}, we compare the values 
of $N(r,k)$ that we obtained in our model to those in the data. 
In spite of the noise in the data, our model is able to reproduce the temporal 
dynamics of $N(r,k)$.
\begin{figure}[htp]
  \begin{center}
    \includegraphics[width=250px]{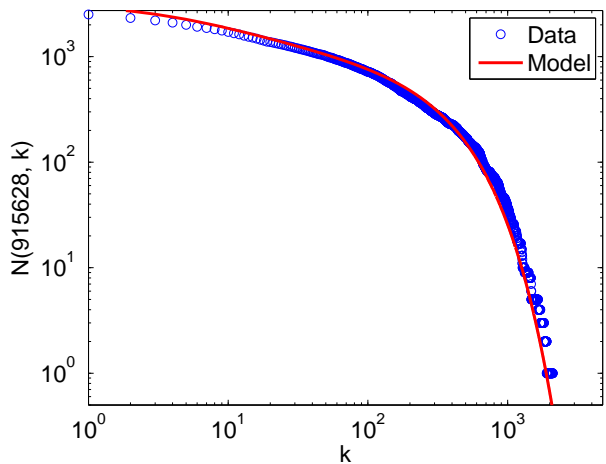}
    \caption{(Color online) Cumulative degree distribution of video nodes
      on the last day (915628 ratings) of the sample from year 2000.  
      We obtained this by solving (\ref{pqNkMasterEqGrwCatNonSat}) using 
      $q = 0.9795$ (red curve) and directly from the data (blue dots).}
    \label{year2000LastDayCumDegDist}
  \end{center}
\end{figure}

In Fig.~\ref{year2000LastDayCumDegDist}, we show the approximation of our model
to the cumulative degree distribution of the videos on the last day of the sample 
(i.e., for all values of $k$ and $r=915628$, the number of ratings at the end of
year 2000), which agrees very well with the data.

Although $q^* \approx 0.9795$ suggests that the way the users choose to rate videos is 
dominated by the popularity of the films, we should stress that the model developed
here is a very simple one. There are probably many other processes influencing the 
decisions of the users, including different external (to the user) factors, such as 
advertisements, press, and the underlying social network the users are embedded in.


\section{Conclusions} \label{SecConclusions}

We have analyzed a large network of video ratings given by 
the users of the Netflix video rental service.  We studied the system using a 
bipartite network of 
videos and users and employed this perspective to reveal interesting features in the 
dynamics of video rating, such as weekly patterns in video ratings and bursts 
of activity followed by long idle periods. We calculated clustering coefficients for 
one-day snapshots, concluding that their low values arise from the presence of 
high-degree nodes (i.e., videos with a large number of ratings and users who rate many 
videos).  We also showed that the degree distributions of both the user and video 
nodes resemble power laws with exponential cutoffs.  

Motivated by the structural and dynamical features we observed in the 
Netflix data, we formulated a mechanism of network evolution in the form of 
``catalog networks'' for bipartite systems.  Such networks are initially empty (aside 
from a seed), and edges are created between two types of nodes based on some 
predefined rules.  New nodes can also be added to the network during the wiring 
process. In our model, we considered a combination of uniform random 
attachment and linear preferential attachment.  We derived a set of coupled 
ordinary differential equations that describe the time-evolution of the degree 
distributions of such catalog networks.  Presupposing this mechanism and employing 
the Netflix data, we found that 
users seem to choose videos according to preferential attachment about 98\% of 
the time and uniform attachment about 2\% of the time.  This suggests that the number 
of ratings for a given video is driven almost completely by its popularity 
(preferential attachment) and only in very small measure by the intrinsic 
preferences of users.  While interesting, the extreme dominance of a 
preferential-attachment mechanism might be due in part to the simplicity of our 
model and the absence of information about the underlying social network 
of the users, which can have considerable influence
over the video choices.
Additionally, our model does not 
incorporate external influences  such as media coverage and promotion campaigns that 
can certainly affect the popularity of 
videos.  One can refine such insights by considering more sophisticated attachment 
mechanisms that incorporate the actual scores of the video ratings (not just their 
existence), the age of the videos, user social networks (see Refs.~\cite{Asur2010} 
and \cite{Ratkiewicz2010} for recent interesting study), interactions among users, 
media presence of videos, and more. Our simple catalog model thereby serves as a 
good starting point for an abundance of interesting generalizations.

The Netflix data, which is both large and publicly available, provides an 
excellent vehicle to study many of the features that have been observed in network 
representations of systems in which agents exercise preferences or choices, such as 
citation, collaboration, and social networks 
\cite{Price:1965, albert-2002-74, Sornette.93.228701, Salganik02102006, Lambiotte2005,
oliveira2005437}.  In this paper, we formulated a catalog model to understand 
the human dynamics of video rating.  In our view, catalog models are suitable in 
many other contexts, 
including studying certain electoral systems (such as the preamble to preferential 
voting elections)~\cite{AustralianPolitics.com}, professional sports drafts 
\cite{HockeyDraft}, and retail shopping. 
To achieve insights in such a diverse array of settings, the catalog model presented 
herein can be generalized in numerous interesting ways to incorporate  
external agents, underlying networks or cliques of individuals, and more.


\section*{Acknowledgements}

We thank M.~Barahona, R.~Desikan, T.~Evans, P.~Ingram, N.~Jones, R.~Lambiotte, 
S.~Lanning, D.~Lazer, P.~Mucha, D.~Plato, S.~Saavedra, D.~Smith, 
and J.~Stark for useful comments and suggestions. We also acknowledge Netflix Inc. 
for providing the data, which was released publicly as part of their prize 
competition. This work was in part done as a dissertation  for the MSc in Mathematical 
Modelling and Scientific Computing at the University of Oxford. MBD was supported 
by a Chevening Scholarship and a 
BBSRC--Microsoft Dorothy Hodgkin Postgraduate Award.  MAP acknowledges 
a research award (\#220020177) from the James S. McDonnell Foundation.  JPO is 
supported by the Fulbright Program.



\begin{thebibliography}{10}

\bibitem{barabasi-2005-435}
A.-L. Barabasi,
\newblock Nature {\bf 435}, 207 (2005).

\bibitem{evans-2008-3}
T.~S. Evans and A.~D.~K. Plato,
\newblock Networks and Heterogeneous Media {\bf 3}, 221 (2008).

\bibitem{J.P.Onnela05012007}
J.-P. Onnela et~al.,
\newblock Proceedings of the National Academy of Sciences {\bf 104}, 7332
  (2007).

\bibitem{albert-2002-74}
R.~Albert and A.-L. Barab\'asi,
\newblock Reviews of Modern Physics {\bf 74}, 47 (2002).

\bibitem{Newman:2003}
M.~E.~J. Newman,
\newblock SIAM Review {\bf 45}, 167 (2003).

\bibitem{NewmanSciCol}
M.~E.~J. Newman,
\newblock Proceedings of the National Academy of Sciences {\bf 98}, 404 (2001)

\bibitem{guido}
G.~Calderelli,
\newblock {\em Scale-Free Networks},
\newblock Oxford University Press, 2007.

\bibitem{HandbookGraphTheory}
J.~L. Gross and J.~Yellen, editors,
\newblock {\em Handbook of Graph Theory},
\newblock CRC Press, 2004.

\bibitem{BarabasiAlbert1999}
A.-L. Barab\'asi and R.~Albert,
\newblock Science {\bf 286}, 509 (1999).

\bibitem{HERBERTA.SIMON12011955}
H.~A. Simon,
\newblock Biometrika {\bf 42}, 425 (1955).

\bibitem{Price:1965}
D.~J. d.~S. Price,
\newblock Science {\bf 149}, 510 (1965).

\bibitem{PhysRevLett.85.4629}
P.~L. Krapivsky, S.~Redner, and F.~Leyvraz,
\newblock Physical Review Letters {\bf 85}, 4629 (2000).

\bibitem{Latapy200831}
M.~Latapy, C.~Magnien, and N.~D. Vecchio,
\newblock Social Networks {\bf 30}, 31  (2008).

\bibitem{MasonPorter05172005}
M.~A. Porter, P.~J. Mucha, M.~E.~J. Newman, and C.~M. Warmbrand,
\newblock Proceedings of the National Academy of Sciences {\bf 102}, 7057
  (2005).

\bibitem{Saavedra07532}
S.~Saavedra, F.~Reed-Tsochas, and B.~Uzzi,
\newblock Nature {\bf 457}, 463 (2009).

\bibitem{Zhang20086869}
P.~Zhang et~al.,
\newblock Physica A {\bf 387}, 6869  (2008).

\bibitem{Guillaume04bipartitegraphs}
J.-L. Guillaume and M.~Latapy,
\newblock Bipartite graphs as models of complex networks,
\newblock in {\em Aspects of Networking}, pages 127--139, Springer, 2004.

\bibitem{PhysRevE.72.036120}
J.~Ohkubo, K.~Tanaka, and T.~Horiguchi,
\newblock Physical Review E {\bf 72}, 036120 (2005).

\bibitem{evans:056101}
T.~S. Evans and A.~D.~K. Plato,
\newblock Physical Review E {\bf 75}, 056101 (2007).

\bibitem{baseball}
S. Saavedra et~al.,
\newblock Physica A {\bf 389}, 1131 (2010).

\bibitem{ceo}
G.~F. Davis, M.~Yoo, and W.~E. Baker,
\newblock Strategic Organization {\bf 1}, 301 (2003).

\bibitem{WattsStrogatz}
D.~J. Watts and S.~H. Strogatz,
\newblock Nature {\bf 393}, 440 (1998).

\bibitem{Dorogovtsev2003396}
S.~N. Dorogovtsev, J.~F.~F. Mendes, and A.~N. Samukhin,
\newblock Nuclear Physics B {\bf 666}, 396  (2003).

\bibitem{PhysRevE.72.026131}
K.~Park, Y.-C. Lai, and N.~Ye,
\newblock Physical Review E {\bf 72}, 026131 (2005).

\bibitem{fan026103}
H.~Fan, Z.~Wang, T.~Ohnishi, H.~Saito, and K.~Aihara,
\newblock Physical Review E {\bf 78}, 026103 (2008).

\bibitem{lind}
J.~Lindquist, J.~Ma, P.~van~den Driessche, and F.~H. Willeboordse,
\newblock Physica D {\bf 238}, 370  (2009).

\bibitem{polya}
G.~Polya,
\newblock Annalea de 1'Institut Henri Poincar\'{e} {\bf 1}, 117 (1931).

\bibitem{feller}
W.~Feller,
\newblock {\em An Introduction to Probability Theory and Its Applications, Vol.
  1},
\newblock John Wiley \& Sons, Inc., 1957.

\bibitem{Godreche0953}
C.~Godr\`{e}che and J.~M. Luck,
\newblock Journal of Physics: Condensed Matter {\bf 14}, 1601 (2002).

\bibitem{Stauffer2007835}
D.~Stauffer, X.~Castell\'o, V.~M. Egu\'iluz, and M.~S. Miguel,
\newblock Physica A {\bf 374}, 835  (2007).

\bibitem{Bentley2003}
R.~A. Bentley and S.~J. Shennan,
\newblock American Antiquity {\bf 68}, 459 (2003).

\bibitem{Netflixprize}
Netflix-Prize,
\newblock \url{http://www.netflixprize.com}.

\bibitem{imdb}
J.~Lorenz,
\newblock European Physical Journal B {\bf 71}, 251 (2009).


\bibitem{netflixPaper}
J.~Bennett and S.~Lanning,
\newblock Proceedings of KDD Cup and Workshop 2007  (2007).

\bibitem{nature06958}
M.~C. Gonz\'alez, C.~A. Hidalgo, and A.-L. Barab\'asi,
\newblock Nature {\bf 453}, 779 (2008).

\bibitem{Sornette.93.228701}
D.~Sornette, F.~Desch\^atres, T.~Gilbert, and Y.~Ageon,
\newblock Physical Review Letters {\bf 93}, 228701 (2004).

\bibitem{oliveira2005437}
J.~G. Oliveira and A.-L. Barab\'asi,
\newblock Nature {\bf 437}, 1251 (2005).

\bibitem{burstbook}
A.-L. Barab\'asi, {\em Bursts: The Hidden Pattern Behind Everything We Do}, 
Dutton Adult, 2010.

\bibitem{Clauset:2007p5520}
A.~Clauset, C.~R. Shalizi, and M.~E.~J. Newman,
\newblock SIAM Review {\bf 51}, 661 (2009).

\bibitem{martacluster}
P.~G. Lind, M.~C. Gonz\'alez, and H.~J. Herrmann,
\newblock Physical Review E {\bf 72}, 056127 (2005).

\bibitem{PhysRevE.71.065103}
J.-P. Onnela, J.~Saram\"aki, J.~Kert\'esz, and K.~Kaski,
\newblock Physical Review E {\bf 71}, 065103 (2005).


\bibitem{AustralianPolitics.com}
\url{australianpolitics.com},
\newblock History \& features of the australian electoral system, 2008,
\newblock Consulted on Feb, 10, 2009.

\bibitem{PhysRevLett.86.3200}
R.~Pastor-Satorras and A.~Vespignani,
\newblock Physical Review Letters {\bf 86}, 3200 (2001).

\bibitem{PhysRevE.71.036127}
A.~Barrat and R.~Pastor-Satorras,
\newblock Physical Review E {\bf 71}, 036127 (2005).

\bibitem{PhysRevE.79.050104}
S.~Johnson, J.~J. Torres, and J.~Marro,
\newblock Physical Review E {\bf 79}, 050104 (2009).

\bibitem{lagarias:112}
J.~C. Lagarias, J.~A. Reeds, M.~H. Wright, and P.~E. Wright,
\newblock SIAM Journal on Optimization {\bf 9}, 112 (1998).


\bibitem{Asur2010}
S.~{Asur} and B.~A. {Huberman},
\newblock (2010),
\newblock arXiv:1003.5699.

\bibitem{Ratkiewicz2010}
J.~Ratkiewicz, F.~Menczer, S.~Fortunato, A.~Flammini, and A.~Vespignani,
\newblock (2010),
\newblock arXiv:1005.2704v1.

\bibitem{Salganik02102006}
M.~J. Salganik, P.~S. Dodds, and D.~J. Watts,
\newblock Science {\bf 311}, 854 (2006).

\bibitem{Lambiotte2005}
R.~Lambiotte and M.~Ausloos,
\newblock Phys. Rev. E {\bf 72}, 066107 (2005).

\bibitem{HockeyDraft}
A.~Summers, T.~Swartz, and R.~Lockhart,
\newblock {\em In Statistical Thinking in Sports}, chapter 15. Optimal drafting
  in hockey pools, pages 249--262,
\newblock Chapman \& Hall/CRC, 2007.

\end{thebibliography}

\bibliographystyle{aip}

\end{document}